\documentclass[conference]{IEEEtran}
\IEEEoverridecommandlockouts
\usepackage{cite}
\usepackage{amsmath,amssymb,amsfonts}
\usepackage{algorithmic}
\usepackage{graphicx}
\usepackage{textcomp}
\usepackage[binary-units]{siunitx}
\newcommand{\CLASSINPUTtoptextmargin}{0.75in}
\newcommand{\CLASSINPUTbottomtextmargin}{1in}

\def\BibTeX{{\rm B\kern-.05em{\sc i\kern-.025em b}\kern-.08em
    T\kern-.1667em\lower.7ex\hbox{E}\kern-.125emX}}

\newcommand{\C}{\mathbb{C}}

\renewcommand{\d}[1]{d#1}

\newcommand{\e}{e}
\renewcommand{\j}{j}

\newcommand{\vct}[1]{\boldsymbol{#1}}
\newcommand{\mtx}[1]{\boldsymbol{#1}}

\newcommand{\<}{\langle}
\renewcommand{\>}{\rangle}
\renewcommand{\H}{\mathrm{H}}

\newcommand{\Span}{\operatorname{Span}}

\newcommand{\set}[1]{\mathcal{#1}}

\DeclareMathOperator*{\argmin}{\text{arg~min}}

\newcommand{\va}{\vct{a}}

\newcommand{\vc}{\vct{c}}

\newcommand{\vu}{\vct{u}}
\newcommand{\vv}{\vct{v}}
\newcommand{\vw}{\vct{w}}

\newcommand{\vy}{\vct{y}}

\newcommand{\valpha}{\vct{\alpha}}

\newcommand{\mA}{\mtx{A}}

\newcommand{\mC}{\mtx{C}}

\newcommand{\mR}{\mtx{R}}

\newcommand{\mU}{\mtx{U}}
\newcommand{\mV}{\mtx{V}}
\newcommand{\mW}{\mtx{W}}

\newcommand{\setH}{\set{H}}

\newcommand{\setW}{\set{W}}

\begin{document}


\title{Robust Broadband Beamforming \\ using Bilinear Programming}

\author{
    \IEEEauthorblockN{Nakul Singh, Coleman DeLude, Mark A.\ Davenport, and Justin Romberg
    \thanks{Email: nsingh360@gatech.edu, cdelude3@gatech.edu, mdav@gatech.edu, jrom@ece.gatech.edu.  This work was supported in part by COGNISENSE, one of seven centers in JUMP 2.0, a Semiconductor Research Corporation (SRC) program sponsored by DARPA, and a grant from Lockheed Martin.}
    }
    \IEEEauthorblockA{
        \textit{School of Electrical and Computer Engineering} \\
        \textit{Georgia Institute of Technology}
    }
}

\maketitle
\thispagestyle{plain}

\begin{abstract}
    We introduce a new method for robust beamforming, where the goal is to estimate a signal from array samples when there is uncertainty in the angle of arrival. Our method offers state-of-the-art performance on narrowband signals and is naturally applied to broadband signals.  Our beamformer operates by treating the forward model for the array samples as unknown. We show that the ``true'' forward model lies in the linear span of a small number of fixed linear systems.  As a result, we can estimate the forward operator and the signal simultaneously by solving a bilinear inverse problem using least squares.  Our numerical experiments show that if the angle of arrival is known to only be within an interval of reasonable size, there is very little loss in estimation performance compared to the case where the angle is known exactly.
\end{abstract}


\section{Introduction}
\label{introduction}

Beamforming, the process of forming an estimate of a signal coming from a fixed direction $\theta$ from the outputs of a multi-sensor array, relies critically on knowing this direction of arrival $\theta$.  When the estimator is ``misaligned,'' i.e.,  using a $\theta$ that is different from the true angle of arrival $\theta_{\mathrm{true}}$, the quality of the estimate can degrade quickly as $\theta$ moves away from $\theta_{\mathrm{true}}$.  This problem is especially pronounced in modern arrays with a large number of elements; large apertures allow us to create tapered beams, but they are also more fragile to misalignment.

While there is a significant amount of literature on making beamformers ``robust'' to unknown $\theta$ (a very nice overview of current methods can be found in \cite{Elbir_2023}),  the most effective approaches known to date 1) apply only to narrowband signals, 2) require solving costly optimization programs to compute weight vectors.  In this paper, we introduce a relatively simple approach based on bilinear programming that naturally includes the broadband case while not being significantly more computationally intensive than broadband beamforming with a known direction of arrival $\theta$.

Recent work in \cite{delude2023slepian} has shown how broadband beamforming can effectively be treated as a linear inverse problem from measurements $\vy\approx\mA(\theta)\valpha$ that can be solved with least squares (this is discussed in more detail in Section~\ref{sec:slepianbf}).  The forward model $\mA(\theta)$ of course depends on the angle of arrival (AOA) $\theta$.  In this paper, we show how we can make this broadband beamforming method robust to uncertainty in $\theta$.  We show how each $A(\theta)$ in a region of angles $\theta\in\Theta$ can be written as a superposition of a relatively small number of fixed matrices that can be pre-computed.  The result is a bilinear least squares problem.  This type of problem is nonconvex, but is prevalent enough in applied mathematics that many different heuristics exist for efficient solvers.  
We demonstrate the effectiveness of one such heuristic, alternating minimization, in the numerical experiments in Section~\ref{sec:results}.  
These results show that the performance penalty for not knowing the angle of arrival exactly is small, and the computation required to achieve this performance is not significantly more than a standard least squares solver.  

In addition to broadband beamforming, the method is also applicable to narrowband beamforming, where there is a well-established literature (some of which is briefly reviewed in Section~\ref{sec:results}).  We show in Section~\ref{sec:results} that our bilinear robust beamformer also offers state-of-the-art performance for narrowband signals.

\section{Formulation}

In this section, we develop the system model for recovering the signal of interest when the arrival angle is unknown. We first present the system model developed in \cite{delude2023slepian} for the case when true AOA is known and then propose a bilinear system model for the cases where we only have a rough estimate of the AOA.

\subsection{Known AOA}
\label{sec:slepianbf}

We consider the beamforming problem where a plane wave signal strikes an $M$ element array at an angle\footnote{This can either represent a standard angle or spherical angle depending on the array configuration. The mathematical formulation is the same in either case.} $\theta$, and $N$ snapshots are recorded, providing a total of $MN$ observations.  
We model the signal of interest $s(t)$ as being bandlimited to $\Omega$, e.g., the spectral support of $s(t)$ is non-zero for $f \in [-\Omega,\Omega]$. 
Over the temporal extent of the $N$ snapshots, we can write $s(t)$ as the superposition of $K$ basis functions $\psi_1(t),\ldots,\psi_K(t)$
\[
    s(t) = \sum_{k=1}^K\alpha_k\psi_k(t).
\]
\topskip=15pt
A natural choice for the $\{\psi_k\}$ are the Slepian basis functions, which provide a compact representation for bandlimited signals supported over a finite interval of time (see \cite{delude2023slepian} for further discussion). Prior to propagation, it is assumed the signal is modulated to a carrier frequency $f_c$ to form $s_\text{mod}(t) = \e^{\j 2\pi f_c t}s(t)$.

The received $s_\text{mod}(t)$ are then demodulated at each array element, reducing the carrier to only a phase offset. We model the array measurements as samples of this demodulated signal. The $n^\text{th}$ sample at the $m^\text{th}$ array element is written as
\begin{equation}
    \label{eq:conv_beamform}
    y_m[n] = \e^{-\j 2\pi f_{c}\tau_{m}(\theta)}\sum_{k=1}^{K}\alpha_k\psi_k(t_n - \tau_m(\theta)) + \textnormal{noise}.\\
\end{equation}
The $\tau_m(\theta)$ in \eqref{eq:conv_beamform} is the delay (relative to array center) of element $m$ for a plane wave arriving from angle $\theta$; $t_n$ denotes the sample time for the $n^\text{th}$ snapshot.  Therefore, a forward model for a single snapshot is given by
\begin{equation}\label{eq:forward_model_1}
    \begin{gathered}
        \vy[n] = \mA_n(\theta)\boldsymbol{\alpha} + \textnormal{noise},\\
    \end{gathered}
\end{equation}
where $\mA_n(\theta)\in\mathbb{C}^{M \times K}$ is the matrix with entries given by $A_n(\theta)[m,k] = e^{-j2\pi  f_{c}\tau_{m}(\theta)}\psi_k(t_n - \tau_m(\theta))$. Stacking the model in \eqref{eq:forward_model_1} for the $N$ snapshots, we arrive at the system model for the array beamforming when the AOA is known:
\begin{equation}\label{eq:system_model_stacked_1}
    \begin{gathered}
        \underbrace{\begin{bmatrix}
            \vy[1]\\
            \vdots\\
            \vy[N]
        \end{bmatrix}}_{\vy} = \underbrace{\begin{bmatrix}
            \mA_1(\theta)\\
            \vdots\\
            \mA_N(\theta)
        \end{bmatrix}}_{\mA(\theta)}\boldsymbol{\alpha} + \textnormal{noise}.
    \end{gathered}
\end{equation}
To recover the Slepian weights ($\boldsymbol{\alpha}$) from \eqref{eq:system_model_stacked_1}, we use least squares which have a closed-form solution:
\begin{equation}\label{eq:linear_least_squares}
    \begin{gathered}
        \Hat{\valpha} = \argmin_{\boldsymbol{\alpha}}\vert\vert \vy - \mathbf{A(\theta)}\boldsymbol{\alpha}\vert\vert_2^2\\
        \implies\hat{\boldsymbol{\alpha}} = (\mathbf{A(\theta)}^H\mathbf{A(\theta)})^{-1}\mathbf{A(\theta)}^H\vy.
    \end{gathered}
\end{equation}

The Slepian beamforming in \eqref{eq:linear_least_squares} works well only when the angle $\theta$ is known; we show how the performance suffers when the estimated angle deviates from the true angle in our numerical experiments in Section~\ref{sec:results} below (in particular, see Figures~\ref{fig:ula_snr} and \ref{fig:upa_snr}). 

It is noteworthy to mention that the dimension $K$ of the low-dimensional subspace spanned by the Slepian basis depends on the true angle of arrival as $K = \lceil 2\Omega T(\theta)\rceil$ where $T(\theta)$ is the array aperture. However, for reasonable deviations in the angle of arrival, the variation in $K$ is insignificant; therefore, in subsequent sections, we will assume the dimension $K$ to be virtually known.

\subsection{Bilinear formulation for unknown AOA}

When $\theta$ is unknown, there is uncertainty in the forward model \eqref{eq:system_model_stacked_1} for the array samples.  We approach this problem by finding a linear embedding for the rows of the $\mA_n(\theta)$ for an interval $\Theta$ of possible AOAs around a fixed angle $\theta_0$: $\Theta = [\theta_0-\Delta,\theta_0+\Delta]$.

To illustrate how this works, let $\va_{\ell}(\theta)\in\C^{K}$ be the $\ell^\text{th}$ row of $\mA(\theta)$.  Note that each value of  $\ell=1,\dots, MN$ has a one-to-one correspondence to a pair $(m,n)$.    As $\theta$ changes, $\va_{\ell}(\theta)$ also changes, but does so smoothly.  The key realization is that the collection of vectors $\{\va_{\ell}(\theta),~\theta\in\Theta\}$ can all be very closely approximated by vectors from a low dimensional subspace.  That is, there exists a $K\times P$ matrix $\mU_\ell$, for a very small $P$, such that for every fixed $\theta\in\Theta$ we have $\va_\ell(\theta) \approx \mU_\ell\vc$ for a short vector $\vc\in\C^P$.  As such, we can write $y_\ell = \vc^\H\mU_\ell^\H\valpha$ where both $\vc$ and $\valpha$ are unknown --- $\valpha$ depends on the signal $s(t)$ and $\vc$ depends on its AOA.

We formalize this carefully as follows.
Define the functions
\[
	w_{\ell,k}(\theta) = e^{-j 2\pi f_{c}\tau_{m}(\theta)}\psi_k(t_{n} - \tau_m(\theta)), ~\theta\in\Theta,
\]
for an interval $\Theta$ as above, and collect these into
\[
	\va_\ell(\theta) = \begin{bmatrix}
		w_{\ell,1}(\theta) \\ \vdots \\ w_{\ell,K}(\theta)
	\end{bmatrix}.
\]
Our problem is that we observe
\begin{equation}
	\label{eq:ym}
	y_\ell = \va_\ell(\theta^*)^\H\valpha + \mathrm{noise},
	\quad \ell=1,\ldots,MN,
\end{equation}
for some unknown $\theta^*\in\Theta$.  We will show how this can be turned into a well-posed bilinear problem by using some basic facts about reproducing kernel Hilbert spaces.

Let $\setH$ be the Hilbert space of functions on $\theta\in\Theta$ generated from the functions $w_{\ell,k}(\theta)$
\begin{equation}
    \begin{aligned}
        \setH = \Span(\{w_{\ell,k}(\theta),~\ell=1,\dots,MN; k=1,\ldots K\}).
    \end{aligned}
\end{equation}
This is a finite-dimensional Hilbert space of continuous functions, and thus it will have a reproducing kernel for any inner product $\<\cdot,\cdot\>$ that we endow it with.  $\setH$ is essentially a Slepian space, as it is generated from modulated versions of small segments of the (approximately) bandlimited functions $\psi_1(t),\ldots,\psi_K(t)$.

Let $\boldsymbol{\setW}_\ell:\setH\rightarrow\C^K$ be the linear operator defined by

\begin{equation}\label{eq:lin_op}
    \boldsymbol{\setW}_\ell x = \begin{bmatrix}
		\left\<x,w_{\ell,1}\right\> \\ \vdots \\ \left\<x,w_{\ell,K}\right\>
	\end{bmatrix}
\end{equation}
and $\boldsymbol{\setW}:\setH\rightarrow\C^{MNK}$ be the concatenation of these operators across the array elements and snapshots:
\[
	\boldsymbol{\setW}x = \begin{bmatrix}
		\boldsymbol{\setW}_1x \\ \vdots \\ \boldsymbol{\setW}_{MN}x
		\end{bmatrix}.
\]
This operator is effectively low rank due to the exponential decay of eigenvalues of the integral operator of which the Slepian basis are the eigenfunctions; for a more detailed analysis, refer \cite{1454379}.  By the spectral theorem, we know that there are (orthonormal) functions $v_1,v_2,\ldots\in\setH$ such that
\begin{align*}
    (\boldsymbol{\setW}^H\boldsymbol{\setW})(\theta,\theta') &= \sum_{\ell,k}w_{\ell,k}(\theta)^Hw_{\ell,k}(\theta')\\ 
    &= \sum_{p=1}^{MNK}\sigma_p^2v_p(\theta)^Hv_p(\theta').
\end{align*}
As noted above, the $w_{\ell,k}(\theta)$ are all just segments of bandlimited signals over small intervals, we can truncate the sum above at some modest $P$:
\[
	(\boldsymbol{\setW}^H\boldsymbol{\setW})(\theta,\theta') \approx \sum_{p=1}^P\sigma_p^2v_p(\theta)^Hv_p(\theta').
\]
Likewise, if we set
\begin{align*}
	\vw(\theta) = \begin{bmatrix}
		\va_1(\theta) \\ \vdots \\ \va_{MN}(\theta) 
	\end{bmatrix},
\end{align*}
\topskip=20pt
we know that the eigenvalue decomposition for the {\small$MNK\times MNK$} matrix can be similarly truncated\footnote{A more detailed analysis of the operator is presented in appendix \ref{appendix:operator}.},
\begin{align*}
    \boldsymbol{\setW}\boldsymbol{\setW}^H &= \int_{\theta\in\Theta}\vw(\theta)\vw(\theta)^\H~\d{\theta}\\
	&= \sum_{p=1}^{MNK}\sigma_p^2\vu_p\vu_p^\H
	\approx \sum_{p=1}^{P}\sigma_p^2\vu_p\vu_p^\H.
\end{align*}
We can thus write for any $x\in\setH$
\[
	\boldsymbol{\setW}x \approx \sum_{p=1}^P\sigma_p\vu_p\<x,v_p\>,
\]
and
\[
	\boldsymbol{\setW}_\ell x \approx \sum_{p=1}^P\sigma_p\vu_{p,\ell}\<x,v_p\>,
\]
where $\vu_{p,\ell}\in\C^K$ is just the ``$\ell^\text{th}$ part'' in $\vu_p\in\C^{MNK}$:
\[
	\vu_p  = \begin{bmatrix} \vu_{p,1} \\ \vdots \\ \vu_{p,MN}
		\end{bmatrix}.
\]
Note that the $\vu_{p,\ell}$ \emph{are not} the left singular vectors of the operator $\boldsymbol{\setW}_\ell$; they are parts of the singular vectors of the joint operator $\boldsymbol{\setW}$.

Returning to our forward model \eqref{eq:ym}, let $\beta^*\in\setH$ be the slice of the reproducing kernel that takes samples at location $\theta^*$.  Then $\va_\ell(\theta^*) = \boldsymbol{\setW}_\ell\beta^*$ and so (ignoring the noise)
\begin{align*}
	y_\ell = \va_{\ell}(\theta^*)^\H\valpha &= \left(\boldsymbol{\setW}_{\ell}\beta^*\right)^\H\valpha \\ 
	&\approx \sum_{p=1}^P\sigma_p\<\beta^*,v_p\>\vu_{p,\ell}^\H\valpha \\
	&= \sum_{p=1}^P c_p\vu_{p,\ell}^\H\valpha \\
	&= \left(\mU_\ell\vc\right)^\H\valpha,
\end{align*}
where $c_p = \sigma_p\<\beta^*,v_p\>$ and the $K\times P$ matrix $\mU_\ell$ is
\begin{align*}
    \mU_\ell = \begin{bmatrix}
		\vu_{1,\ell} & \ldots &\vu_{P,\ell}\\
	\end{bmatrix}.
\end{align*}
\emph{Note that $\vc$ is independent of $l$}.  Thus our bilinear problem is to recover $\vc\in\C^P$ and $\alpha\in\C^K$ from measurements $y_\ell$ given the forward model

\begin{equation}\label{eq:bilinear_scalar}
    y_\ell = \vc^\H(\mU_\ell)^\H\valpha ~+~ \mathrm{noise},
\end{equation}
where the $\mU_\ell$ are known $K\times P$ matrices that can be computed beforehand. Ideally, $\mU_\ell$ are calculated after taking the singular value decomposition of the matrix $\boldsymbol{\setW}\boldsymbol{\setW}^H$. However, for a large array with many snapshots, performing the SVD of this matrix can be computationally expensive. To circumvent this issue, we approximate the operator $\boldsymbol{\setW}$ as a matrix that allows faster computation of the SVD. Instead of defining the functions $w_{\ell,k}(\theta)$ over a continuous interval $\Theta$, we discretize the interval by uniformly sampling $Q$ points $\{\theta_{1},\hdots,\theta_{Q}\} \in \Theta$. Corresponding to these sampled points, we define $\mW_\ell:\C^{Q}\rightarrow\C^K$ as
\begin{align*}
    \mW_\ell = \begin{bmatrix}
        w_{\ell,1}(\theta_1)&&\ldots &&w_{\ell,1}(\theta_Q)\\
        \vdots && \ddots && \vdots\\
        w_{\ell,K}(\theta_1)&&\ldots&& w_{\ell,K}(\theta_Q)\\
    \end{bmatrix},
\end{align*}
and $\mW:\C^{Q}\rightarrow\C^{MNK}$ becomes the concatenation of these approximate operators.
We can now evaluate the SVD of the {\small$Q\times Q$} matrix $\mW^H\mW$ much faster since {\small$Q \ll MNK$} for large arrays. Using the sampled orthonormal functions $\vv_{p}$ we get $\vu_{p} = \mW\vv_{p}/\sigma_{p}$ which are then used to compute $\mU_\ell$. 

The next section describes the alternating minimization method used to solve the bilinear least squares resulting from the model devised in \eqref{eq:bilinear_scalar}.

\section{Alternating minimization for bilinear least squares}

To perform alternating minimization, we first stack the bilinear formulation derived in \eqref{eq:bilinear_scalar} across array elements and snapshots to arrive at the following forward model, 

\begin{equation*}\label{eq:bilinear_model_stacked_1}
    \begin{gathered}
        \underbrace{\begin{bmatrix}
            y_1\\
            \vdots\\
            y_{MN}
        \end{bmatrix}}_{\vy} =  \mC\underbrace{\begin{bmatrix}
            \mU_1^H\\
            \vdots\\
            \mU^H_{MN}
        \end{bmatrix}}_{\Bar{\mU}}\valpha + \text{noise},\\
        \mC\in\mathbb{C}^{MN\times PMN}\text{ s.t. }C[i,j] = \begin{cases}
            [c^H]_j \text{ }(i-1)P\leq j \leq iP\\
            0 \text{ otherwise}.
        \end{cases}
    \end{gathered}
\end{equation*} 

A similar model with block structure over $\valpha$ instead of $\vc$ is given by,

\begin{equation*}\label{eq:bilinear_model_stacked_2}
    \begin{gathered}
        \underbrace{\begin{bmatrix}
            (y_1)^*\\
            \vdots\\
            (y_{MN})^*
        \end{bmatrix}}_{\Bar{\vy}} =  \mV\underbrace{\begin{bmatrix}
            \mU_1\\
            \vdots\\
            \mU_{MN}
        \end{bmatrix}}_{\mU}\vc + \text{noise},\\
        \mV\in\mathbb{C}^{MN\times KMN}\text{ s.t. }V[i,j] = \begin{cases}
            [\alpha^H]_j \text{ }(i-1)K\leq j \leq iK\\
            0 \text{ otherwise}.
        \end{cases}
    \end{gathered}
\end{equation*}

Using these models, we have two equivalent bilinear least squares problems that can be solved alternately while fixing one of the variables. Keeping $\vc^t$ fixed at the $t^\text{th}$ iteration, we obtain a linear least squares problem in $\valpha$:
\begin{equation*}
    \begin{gathered}
        \Hat{\valpha}^{t+1} = \argmin_{\valpha}\vert\vert \vy - \underbrace{\mC^t\Bar{\mU}}_{\mA^t_{\valpha}}\valpha\vert\vert_2^2\\
        \implies \Hat{\valpha}^{t+1} = ((\mA^t_{\valpha})^H\mA^t_{\valpha})^{-1}(\mA^t_{\valpha})^H\vy.
    \end{gathered}
\end{equation*}
Next fixing $\valpha^{t+1}$ we obtain a linear least squares problem in $\vc$:
\begin{align*}
    \begin{gathered}
        \Hat{\vc}^{t+2} = \argmin_{\vc}\vert\vert \Bar{\vy} - \underbrace{\mV^{t+1}\mU}_{\mA^{t+1}_{\vc}}\vc\vert\vert_2^2\\
        \implies \Hat{\vc}^{t+2} = ((\mA^{t+1}_{\vc})^H\mA^{t+1}_{\vc})^{-1}(\mA^{t+1}_{\vc})^H\Bar{\vy}.
    \end{gathered}
\end{align*}
It is important to note that we solve an unconstrained bilinear least squares problem. Therefore, the optimizers of the program are not unique, i.e., if the algorithm recovers weights $\hat{\valpha}$, $\hat{\vc}$ then $\hat{\valpha}/\gamma$ and $\gamma\hat{\vc}$ are also feasible solutions of the problem for $\gamma \neq 0$. To find $\gamma$ associated with the true angle, we can create a sequence of weights $\vc_{q}$ for each $\theta_{q}$ sampled from $\Theta$. For a sufficiently fine-grained sampling, the true $\vc$ will be well-approximated by some $ \vc \in \{\vc_1,\hdots,\vc_Q\}$. Thus, if we let 
\begin{equation*}
        q^{*} = \argmin_{q}\Big\vert\Big\vert \vc_{q} - \gamma_{q}\hat{\vc}\Big\vert\Big\vert^2_2 
\end{equation*}
where $\gamma_{q} = \frac{\vc_{q}^T\hat{\vc}}{\hat{\vc}^T\hat{\vc}}$, then $\gamma_{q^*}$ is the offset constant linked with the true angle and $\Hat{\valpha}/\gamma_{q*}$ provides the estimate of the true Slepian weights. Other approaches, like line spectral estimation \cite{delude2022iterative}, can also be adapted for fine-tuning the constant $\gamma$. We empirically observe that using a sufficiently good initialization for $\vc^{0}$ results in $\gamma \approx 1$.  Thus, we start the alternating minimization scheme described above with an initial value of $\vc^{0}$ corresponding to the mid-point of the interval $\Theta$, allowing the algorithm to converge in $10$ to $15$ iterations for the experiments illustrated in the following section. 

\section{Numerical results}
\label{sec:results}

We use the linear Slepian model in \eqref{eq:linear_least_squares} to perform two kinds of beamforming to benchmark the performance of the proposed model. In the first scenario, we assume that the angle of arrival is known and true Slepian weights ($\boldsymbol{\alpha}_{true}$) are evaluated using \eqref{eq:linear_least_squares}. In the second case, we use the midpoint of the uncertainty interval ($\Theta$) as an estimate for the angle to recover a crude estimate of the Slepian weights. These are then used for a quantitative comparison against the estimates obtained from the proposed method. In addition, in the broadband regime, we also compare our algorithm against conventional approaches like delay and sum beamforming \cite{book}. In the narrowband regime, we compare the performance with several robust adaptive beamforming methods.

\begin{figure}[h]
	\centering	
	\includegraphics[width=0.42\textwidth]{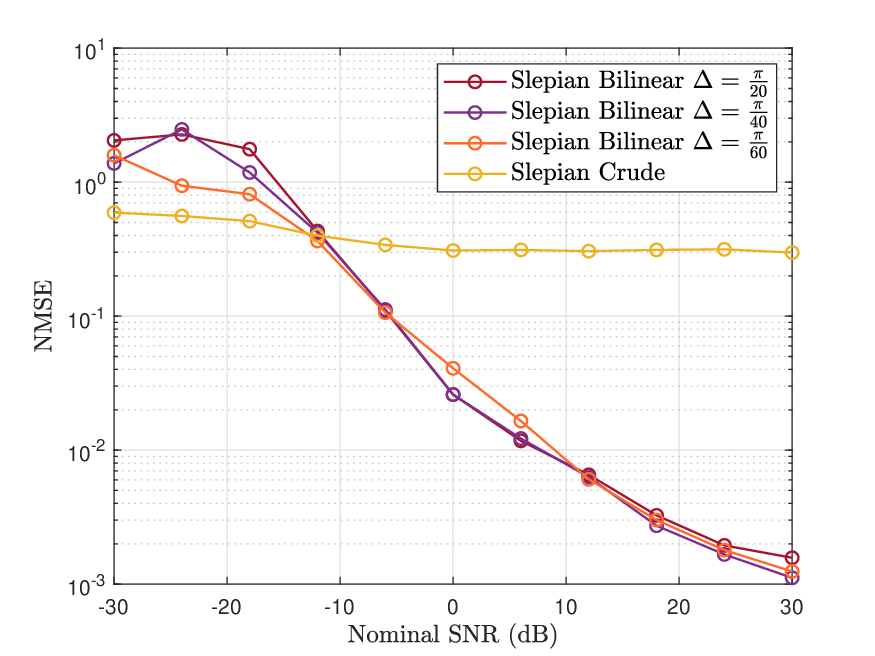}
	\caption{\small \sl NMSE of Slepian coefficients ($\boldsymbol{\alpha}$) v/s nominal SNR for 64-element ULA with 32 snapshots in the broadband regime with bandwidth $\SI{5}{\giga\hertz}$.}
	\label{fig:ula_nmse}
\end{figure}

\begin{figure}[h]
	\centering	\includegraphics[width=0.42\textwidth]{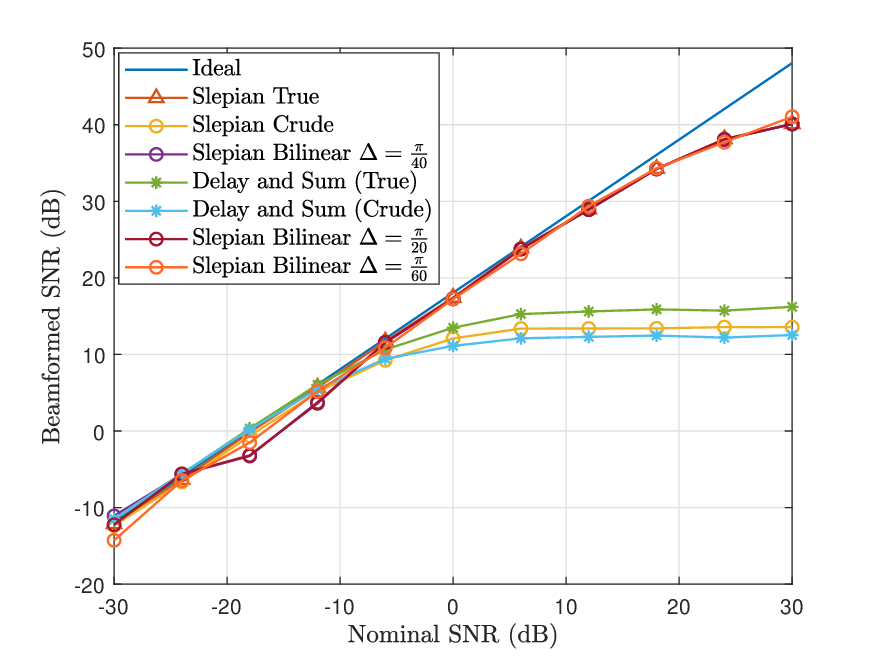}
	\caption{\small \sl Nominal SNR compared to Beamformed SNR for a 64-element ULA with 32 snapshots in the broadband regime with bandwidth $\SI{5}{\giga\hertz}$.}
	\label{fig:ula_snr}
\end{figure}

We perform this experiment on two array geometries, a one-dimensional uniform linear array (ULA) and a two-dimensional uniform planar array (UPA), for the broadband system to validate our findings. In both array designs, the incoming signal follows a sum of sinusoid model corrupted by complex Gaussian noise \cite{DAVENPORT2012438}. We first consider a $64$ element ULA with $N=32$ snapshots and three kind of uncertainty interval $\Theta$, i.e. the interval $\Theta \equiv [\frac{\pi}{6}-\Delta, \frac{\pi}{6} + \Delta]$ where $\Delta\in\{\frac{\pi}{60},\frac{\pi}{40},\frac{\pi}{20}\}$. Fig.\ref{fig:ula_nmse} plots the NMSE $\frac{\vert\vert \boldsymbol{\hat{\alpha}} -\boldsymbol{\alpha}_{true}\vert\vert_2}{\vert\vert\boldsymbol{\alpha_{true}}\vert\vert_2}$ for the crude linear Slepian and the proposed bilinear Slepian, we can observe from Fig.~\ref{fig:ula_nmse} that the proposed method can recover the true Slepian coefficients as the nominal SNR increases for all levels of uncertainty interval. The marginal loss in beamforming performance compared to true linear Slepian confirms this observation as seen in Fig.~\ref{fig:ula_snr}. On the other hand, beamforming done using rough angle estimates suffers badly at higher nominal SNRs due to large deviations from the true coefficients.

\begin{figure}[h]
	\centering	
	\includegraphics[width=0.42\textwidth]{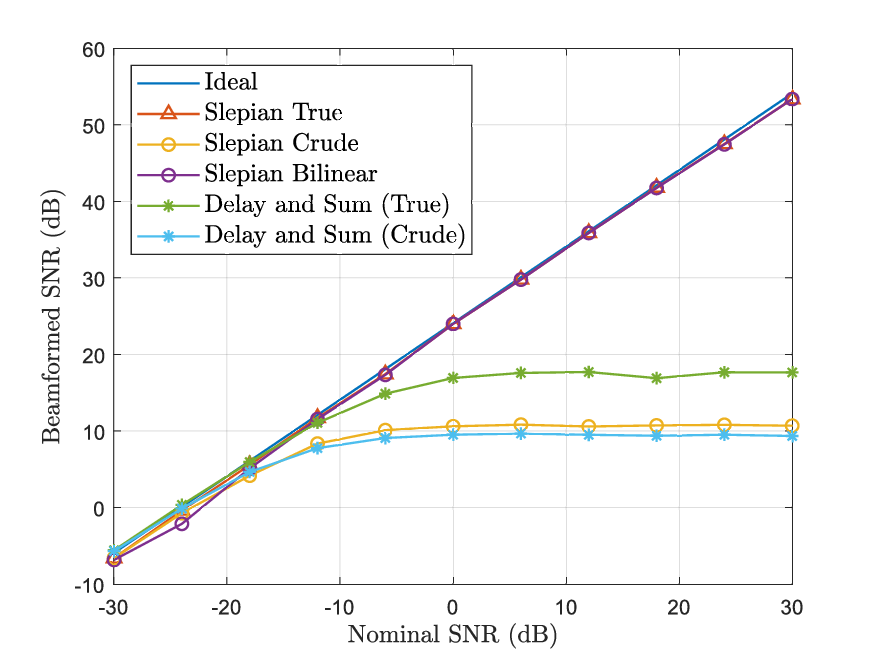}
	\caption{\small \sl Nominal SNR compared to Beamformed SNR for a 16$\times$16 element UPA with 32 snapshots in the broadband regime with bandwidth $\SI{5}{\giga\hertz}$.}
	\label{fig:upa_snr}
\end{figure}

Next, we use a 16$\times$16 element UPA with  $N=32$ snapshots. Two different uncertainty intervals are defined for the elevation $\theta$ and the azimuth angle $\phi$; $\Theta_{\theta}\equiv [\frac{\pi}{4}-\frac{\pi}{40}, \frac{\pi}{4} + \frac{\pi}{40}]$ and \textbf{$\Theta_{\phi}\equiv [\frac{\pi}{3}-\frac{\pi}{40}, \frac{\pi}{3} + \frac{\pi}{40}]$}. Like the ULA experiment, the bilinear method is better at recovering the true Slepian weights than the crude linear model. The delay and sum beamforming employs fractional delay filters consisting of sinc interpolators truncated to $R$ taps. The ideal beamformed SNR is achieved when $R \rightarrow \infty$, for a truncated/finite $R$, the filter bias increases with an increase in SNR. This causes deterioration in beamformed SNR as observed in Fig.~\ref{fig:ula_snr} and Fig.~\ref{fig:upa_snr}. Further degradation is observed in delay and sum when using a crude angle due to an inaccurate estimate of the phase delays.

\begin{figure}[h]
	\centering	
	\includegraphics[width=0.42\textwidth]{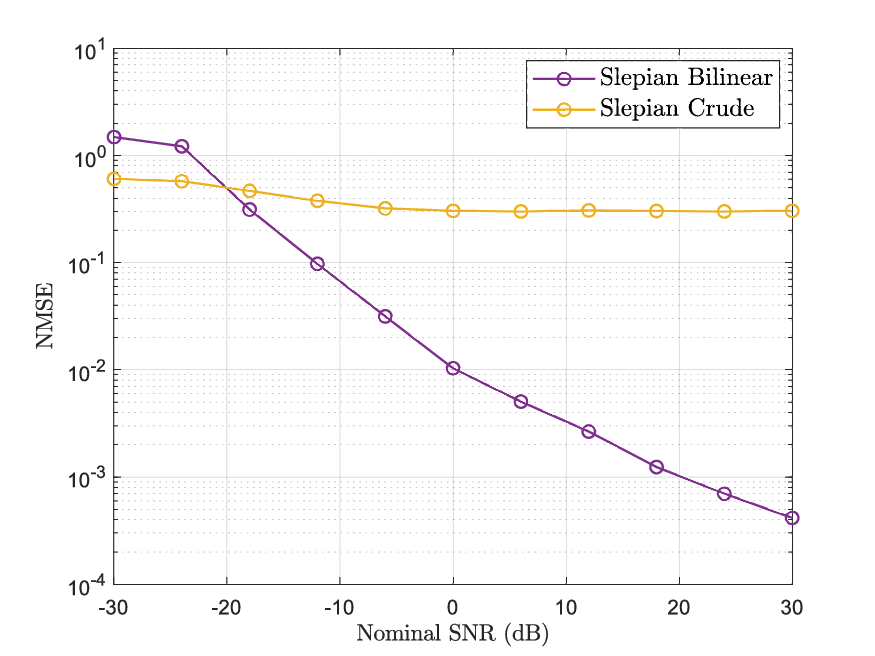}
	\caption{\small \sl NMSE of Slepian coefficients ($\boldsymbol{\alpha}$) v/s nominal SNR for 16$\times$16-element UPA with 32 snapshots in the broadband regime with bandwidth $\SI{5}{\giga\hertz}$.}
	\label{fig:upa_nmse}
\end{figure}
\begin{figure}[h]
	\centering	
	\includegraphics[width=0.42\textwidth]{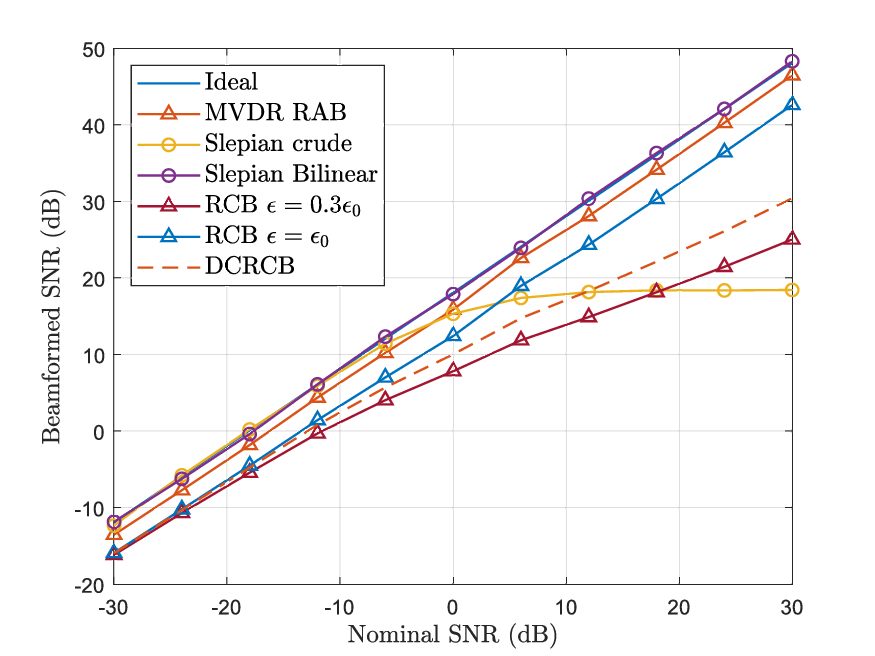}
	\caption{\small \sl Nominal SNR compared to Beamformed SNR for a 64-element ULA with 32 snapshots in the narrowband regime with bandwidth $\SI{5}{\mega\hertz}$.}
	\label{fig:ula_narrowband}
\end{figure}

In the narrowband regime, we compare our algorithm against robust adaptive approaches like RCB \cite{1206680}, DCRCB \cite{1323250}, and MVDR-RAB \cite{6159099}, and the beamformed SNR is plotted in Fig.~\ref{fig:ula_narrowband}. RCB uses a spherical uncertainty set to recover a steering vector estimate. DCRCB adds a norm constraint to the constraints defined by RCB, slightly improving the beamformed SNR. The spherical uncertainty constraint in RCB, DCRCB is sensitive to the choice of parameter $\epsilon$. We observe a performance drop in these algorithms for $\epsilon$ values further away from the true $\epsilon_{0}$. They are also sensitive to array calibration errors and require perfect knowledge of the array geometry and noise characteristics. MVDR-RAB, on the other hand, uses minimum prior information and assumes the angle lies in an angular sector similar to $\Theta$ defined in our case. The constraint added in MVDR-RAB ensures that the estimated steering vector does not converge to any interfering signal outside the interval $\Theta$. The optimization problem solved in MVDR-RAB is given by 
\begin{equation}\label{eq:mvdr_Rab}
    \begin{gathered}
        \argmin_{\va}~ \va^H\mR^{-1}\va~
        \text{s.t. } ~\va^{H}\Tilde{\mC}\va \leq \Delta_{0},~
        \vert\vert \va\vert\vert^2_2 = M,
    \end{gathered}
\end{equation}
where $\Tilde{\mC} = \int_{\Tilde{\Theta}}\va(\theta)\va^H(\theta)d\theta$. In the narrowband system, MVDR-RAB performs as well as the proposed method with only a marginal drop in performance; however, extending it to broadband is not trivial due to several drawbacks inherent in the optimization problem setup. As seen from \eqref{eq:mvdr_Rab}, the norm constraint makes the problem non-convex, requiring SDP-relaxation to find an optimal solution. In the case of broadband, it is difficult to implement SDP relaxation due to the matrices involved. Also, the objective function might not remain convex when extended to broadband since there is no straightforward analog of the steering vector for broadband. In contrast, the proposed bilinear algorithm is simple and works regardless of our operating regime.

\bibliographystyle{IEEEtran}
\bibliography{IEEEexample}

\begin{thebibliography}{10}
\providecommand{\url}[1]{#1}
\csname url@samestyle\endcsname
\providecommand{\newblock}{\relax}
\providecommand{\bibinfo}[2]{#2}
\providecommand{\BIBentrySTDinterwordspacing}{\spaceskip=0pt\relax}
\providecommand{\BIBentryALTinterwordstretchfactor}{4}
\providecommand{\BIBentryALTinterwordspacing}{\spaceskip=\fontdimen2\font plus
\BIBentryALTinterwordstretchfactor\fontdimen3\font minus \fontdimen4\font\relax}
\providecommand{\BIBforeignlanguage}[2]{{%
\expandafter\ifx\csname l@#1\endcsname\relax
\typeout{** WARNING: IEEEtran.bst: No hyphenation pattern has been}%
\typeout{** loaded for the language `#1'. Using the pattern for}%
\typeout{** the default language instead.}%
\else
\language=\csname l@#1\endcsname
\fi
#2}}
\providecommand{\BIBdecl}{\relax}
\BIBdecl

\bibitem{Elbir_2023}
A.~Elbir, K.~Mishra, S.~Vorobyov, and R.~Heath, ``{Twenty-Five Years of Advances in Beamforming: From convex and nonconvex optimization to learning techniques},'' \emph{{IEEE} Signal Process. Mag.}, vol.~40, no.~4, p. 118–131, 2023.

\bibitem{delude2023slepian}
C.~DeLude, M.~Davenport, and J.~Romberg, ``{Slepian Beamforming: Broadband Beamforming using Streaming Least Squares},'' arXiv:2312.03922, December 2023.

\bibitem{1454379}
D.~Slepian, ``On bandwidth,'' \emph{Proceedings of the IEEE}, vol.~64, no.~3, pp. 292--300, 1976.

\bibitem{delude2022iterative}
C.~DeLude, R.~Sharma, S.~Karnik, C.~Hood, M.~Davenport, and J.~Romberg, ``{Iterative Broadband Source Localization},'' \emph{IEEE J. Sel. Areas Inf. Theory}, vol.~4, pp. 453--469, 2023.

\bibitem{book}
H.~V. Trees, \emph{Optimum {A}rray {P}rocessing: {P}art {IV} of {D}etection, {E}stimation, and {M}odulation {T}heory}.\hskip 1em plus 0.5em minus 0.4em\relax Wiley Interscience, 2002.

\bibitem{DAVENPORT2012438}
M.~Davenport and M.~Wakin, ``Compressive sensing of analog signals using {D}iscrete {P}rolate {S}pheroidal {S}equences,'' \emph{Appl. Comput. Harmon. Anal.}, vol.~33, no.~3, pp. 438--472, 2012.

\bibitem{1206680}
J.~Li, P.~Stoica, and Z.~Wang, ``{On robust Capon beamforming and diagonal loading},'' \emph{{IEEE} Trans. Signal Process.}, vol.~51, no.~7, pp. 1702--1715, 2003.

\bibitem{1323250}
------, ``{Doubly constrained robust Capon beamformer},'' \emph{{IEEE} Trans. Signal Process.}, vol.~52, no.~9, pp. 2407--2423, 2004.

\bibitem{6159099}
A.~Khabbazibasmenj, S.~Vorobyov, and A.~Hassanien, ``{Robust Adaptive Beamforming Based on Steering Vector Estimation With as Little as Possible Prior Information},'' \emph{{IEEE} Trans. Signal Process.}, vol.~60, no.~6, pp. 2974--2987, 2012.

\bibitem{weidmann2012li}
J.~Weidmann, \emph{Linear Operators in Hilbert Spaces}.\hskip 1em plus 0.5em minus 0.4em\relax Springer, 2012.

\bibitem{crane2020th}
D.~Crane, ``The singular value expansion for compact and non-compact operators,'' Doctoral Dissertation, Michigan Technological University, 2020.

\bibitem{hutson2005ap}
V.~Hutson, J.~Pym, and M.~Cloud, \emph{Applications of Functional Analysis and Operator Theory}, 2nd~ed.\hskip 1em plus 0.5em minus 0.4em\relax Elsevier, 2005.

\end{thebibliography}

\begin{appendices}
    \section{Singular value expansion for operator $\boldsymbol{\setW}$}
    \label{appendix:operator}
    To begin, recall that the finite-dimensional Hilbert space we are examining is
\begin{align*}
    \setH = \text{Span}\left (\{w_\ell(\theta)\}_{\ell=1}^{L} \right)
\end{align*}
where each basis function $w_\ell(\theta)$ operates over the range $\theta \in \Theta$. We use a slightly different indexing for ease of notation. The operator we are examining, $\boldsymbol{\setW}:\setH\to\mathbb{C}^L$, is given by
\begin{align*}
    \boldsymbol{\setW} x = 
    \begin{bmatrix}
    \<x,w_1\>\\
    \<x,w_2\>\\
    \vdots\\
    \<x,w_L\>
    \end{bmatrix}.
\end{align*}
It is an operator between two finite-dimensional Hilbert spaces and, therefore, is compact \cite[Chapt.\ 4]{weidmann2012li}. The first step in determining the Singular value expansion (SVE) is determining the adjoint of the operator. Note that the inner product for $\setH$ is given by $\<w,x\>_{\setH} = \int_{\Theta}x(\theta)\bar{w}(\theta)d\theta$,a and the innner product for $\mathbb{C}^L$ is the standard vector inner product $\<\cdot,\cdot\>_{\mathbb{C}}$. Then we have
\begin{align*}
    \left\<\boldsymbol{\setW} x,y   \right\>_{\mathbb{C}} &= \sum_{\ell=1}^L y_{\ell} \overline{\<x,w_\ell\>}_{\setH}
    = \sum_{\ell=1}^L y_{\ell} \<w_{\ell},x\>_{\setH}\\
    &= \sum_{\ell=1}^L y_{\ell} \int_{\Theta} w_{\ell}(\theta) \overline{x}(\theta) d\theta\\
    &= \int_{\Theta}\overline{x}(\theta) \sum_{\ell=1}^L y_{\ell}w_{\ell}(\theta) d\theta\\
    &= \left\< x, \sum_{\ell=1}^L y_{\ell}w_{\ell} \right\>_{\setH},
\end{align*}
Hence the adjoint $\boldsymbol{\setW}^H: \mathbb{C}^L \to \setH$ is given by 
\begin{align*}
    \boldsymbol{\setW}^Hy = \sum_{\ell=1}^L y_{\ell} w_{\ell}(\theta).
\end{align*}

Since the adjoint of a compact operator is also compact, therefore the composite operators $\boldsymbol{\setW}\boldsymbol{\setW}^H$ and $\boldsymbol{\setW}^H\boldsymbol{\setW}$ are also compact \cite[Chapt.\ 4]{weidmann2012li}. It is easy to verify they are also self-adjoint and, therefore, admit an eigendecomposition \cite[Chapt. \ 1]{crane2020th}. It is the case that the eigenvectors of $\boldsymbol{\setW}^H\boldsymbol{\setW}$ are the sequence of functions {$v_{\ell}$}, and similarly, the eigenvectors of $\boldsymbol{\setW}\boldsymbol{\setW}^H$ are the sequence of vectors {$\vu_{\ell}$} \cite[Chapt. \ 7 ]{hutson2005ap}. Through similar reasoning, it is also the case that the non-zero eigenvalues of these self-adjoint operators are the same and equal to $\{\sigma^2_{\ell}\}$.
From this point, it is not too difficult to see that
\begin{align*}
    \boldsymbol{\setW}^H\boldsymbol{\setW} x &= \sum_{\ell=1}^L w_\ell(\theta)\<w_{\ell},x\>_{\setH}\\
    & = \sum_{\ell=1}^L \sigma_{\ell}^2 v_\ell(\theta)\<v_{\ell},x\>_{\setH}
\end{align*}
where the second line follows from the spectral theorem \cite{crane2020th}. We can state this more concisely as
\begin{align*}
    \boldsymbol{\setW}^H\boldsymbol{\setW} = \sum_{\ell=1}^L w_\ell(\theta)\otimes w_\ell(\theta') = \sum_{\ell=1}^L \sigma_{\ell}^2 v_\ell(\theta)\otimes v_{\ell}(\theta')
\end{align*}
where the addition of a $\theta'$ is meant to emphasize that the inner product against the rightmost eigenfunction is taken independently from the leftmost. In a similar manner, we can solve for the left eigenfunctions by examining
\begin{align*}
    \boldsymbol{\setW}\boldsymbol{\setW}^Hx &= 
    \begin{bmatrix}
    \<w_1,\sum_{\ell=1}^L x_{\ell} w_{\ell}(\theta)\>\\
    \<w_2,\sum_{\ell=1}^L x_{\ell} w_{\ell}(\theta)\>\\
    \vdots\\
    \<w_L,\sum_{\ell=1}^L y_{\ell} w_{\ell}(\theta)\>
    \end{bmatrix}.\\
    &= \int_{\Theta} \begin{bmatrix}
        w_1(\theta)\overline{w_1}(\theta)  & \dots & w_L(\theta)\overline{w_1}(\theta) \\
        \vdots & \ddots & \vdots\\
         w_1(\theta)\overline{w_L}(\theta)  & \dots &  w_L(\theta)\overline{w_L}(\theta) 
    \end{bmatrix} 
    \begin{bmatrix}
        x_1\\
        x_2\\
        \vdots\\
        x_{L}
    \end{bmatrix}d\theta.
\end{align*}
which is just an $L\times L$ symmetric matrix operator, and as with all such matrices, we can from an eigendecomposition on it to get a set of eigenvectors $\{\vu_{\ell}\}_{\ell=1}^L$. Putting it all together, we have
\begin{align*}
    \boldsymbol{\setW} = \sum_{\ell=1}^L \sigma_{\ell}\vu_{\ell} \otimes v_{\ell}
\end{align*}
such that
\begin{align*}
    \boldsymbol{\setW} x = \sum_{\ell=1}^L \sigma_{\ell}\vu_{\ell} \<x,v_{\ell}\>_{\setH}.
\end{align*}
Thus, the calculation and justification of the factorization present in the robust beamforming method are concluded.
\end{appendices}
\end{document}